\documentclass{optica-article}

\journal{opticajournal} % for journals or Optica Open

\articletype{Research Article}

\usepackage{lineno}
\usepackage{siunitx}
\usepackage{float}
\usepackage{caption}
\usepackage{subcaption}
\usepackage{comment}
\usepackage{fancyhdr}
\usepackage{lastpage}
\usepackage{soul,indentfirst}
\usepackage{bm}
\begin{document}

\title{Polarimetric Backscattering Setup for Quantitative Scattering Parameters Retrieval} 

\author{B. P. Singh\authormark{1}, V. Stefanov\authormark{1,*}, N. A. Coorens\authormark{1,2}, and A. Stefanov\authormark{1}}

\address{\authormark{1}Institute of Applied Physics, University of Bern, Switzerland}
\address{\authormark{2}Faculty of Science and Technology, University of Twente, the Netherlands}

\email{\authormark{*}vladislav.stefanov@unibe.ch}

\begin{abstract*} 
We present two backscattering polarimetric scanning setups based on  point-illumination schemes, that are designed to probe the optical properties of subsurface media. We describe their advantages and limitations, characterize their performance in Mueller matrix determination, and demonstrate their ability to retrieve unknown sample parameters (such as the scattering coefficient and anisotropy factor) using a maximum likelihood approach. As an experimental validation, a polystyrene suspension is investigated as a representative case. By applying Fisher information formalism on the theoretical model, we estimate boundaries of the retrieved parameters.

\end{abstract*}

\section{Introduction}

Polarimetric imaging techniques have found growing applications across various biomedical and material science domains due to their sensitivity to microstructural and optical anisotropies \cite{8744376, louie2021constructing, qi2017mueller, vitkin2015tissue, li2022polarimetric}. In biomedical imaging, these techniques have been used to assess tissue morphology \cite{AHMAD2020101708}, detect early-stage cancerous changes \cite{10.1117/1.JBO.19.7.076013}, and monitor structural alterations in layered or fibrous tissues such as skin \cite{Novikova:16}, brain and muscle \cite{Novikova2023}. By analyzing the polarization changes of light scattered from tissue surfaces and subsurface layers, backscattering polarimetric methods enable contrast mechanisms that are not accessible with conventional intensity-based imaging \cite{qi2017mueller, novikova2016special}. Beyond biomedicine, these techniques are also employed in materials inspection \cite{Deibler:01}, including the evaluation of surface roughness \cite{10.1117/1.OE.62.11.114104}, stress-induced birefringence  and subsurface defects in polymers and composites \cite{10.1117/1.OE.63.9.094104}. Their ability to provide non-contact, label-free, and depth-sensitive information makes them valuable for both clinical research and industrial diagnostics.

Traditional wide-field polarimetry assumes illumination of full sample by a broad light beam, and it is well suited for applications in transparent, semitransparent or thin medium in forwardscattering mode \cite{guo2013study} and for studies of surfaces in backreflection mode \cite{vitkin2015tissue, rodriguez2022polarimetric}. 
However, in backscattering configuration a wide-field illumination typically penetrates only a few hundred microns, the information obtained in backscattering images is superficial and limited to the surface. If the surface is uneven or has residues, the resulting data may be distorted or unreliable. 

Those limitations can be addressed by backscattering scanning polarimetry, where a polarized and focused laser beam is directed onto the sample \cite{Hornung:19, jiao2000depth, ni2004time, hielscher1997diffuse, rakovic1999light}. This focused beam penetrates several millimeters into the tissue, and the backscattered light provides information about the deeper structural and optical properties of the tissue \cite{Stefanov:23}.

However, in such polarimetric setup the deep backscattering signal is much lower than the signal reflected by the surface. For an unbiased interpretation of the measurements, the illumination beam must be normal to the sample surface and well separated from the specularly reflected signal. Here, we compare two configurations that enable both illumination and light collection at normal incidence. The first employs a beamsplitter combined with an occultation mask to block the central area, while the second uses a small $\ang{45}$ mirror that naturally blocks the reflected light.

We present the general characteristics of both setups and discuss their experimental limitations and advantages for data analysis, thereby clarifying the measurement conditions under which each configuration is most suitable. To quantitatively evaluate their performance, we address the retrieval of the optical scattering properties of a reference medium. 

Since conventional sample analysis based on identifying regions with distinct optical properties (such as retardance, diattenuation, depolarization, etc.~\cite{Lu:96}) is hardly suited for data obtained from a single illumination point, we adopt a maximum-likelihood approach to compare the experimental results with theoretically predicted ones. This approach requires a well-characterized model sample (here, a polystyrene suspension) and a sufficiently large reference database generated via Monte Carlo simulations of Mie scattering \cite{Ramella-Roman:051,Ramella-Roman:052}. The consistency between experimental and simulated results therefore provides a good indicator for evaluating the quality of the setups.

This paper is organized as follows. Section~2 presents the mathematical background of the Mueller matrix formalism for the considered setups. Section~3 describes the setup configurations, including technical details of the hardware components and the calibration procedure. Section~4 is dedicated to the experimental characterization of the setups, beginning with a description of the scattering medium model, followed by a numerical assessment of statistical errors based on the Fisher information framework, and concluding with a comparison of the measurements and simulated data. Section~5 summarizes the results and concludes the paper.

\section{Mathematical background}
A typical polarimetric imaging setup consists of two arms. One arm is used to produce different polarization states and shine those states on a tissue sample and another arm is used to record different polarization states of the signal coming from the sample. From the set of all measurements, an intensity matrix is produced, where each element of the matrix corresponds to the input state and its measured intensity with respect to the analyzed state. For linear light-tissue interactions the polarimetric properties can be fully characterized by the Perrin-Mueller matrix (hereafter referred to as the Mueller matrix, MM) \cite{li2022polarimetric}, which can be calculated from intensity matrix, that contains at least 16 independent intensity measurements. The calculated $4\times4$ matrix provides a concise and comprehensive representation of the sample's polarimetric properties. After processing, it enables the extraction of physical characteristics (such as structural and compositional features) of the tissue under investigation from its optical properties (such as scattering, birefringence, diattenuation) \cite{Hornung:19,gil2022polarized,ramella2022polarized}.

Polarization of electromagnetic light is defined by the shape and orientation of the locus of the electric field vector as a function of time. If the electric field vector traces a stationary curve over time, the electromagnetic radiation is said to be polarized \cite{goldstein2017polarized}. There are two main mathematical formalisms used to describe polarized light: (i) Jones calculus \cite{yakovlev2015polarization}, which represents light in terms of the amplitude of the electric field components, and (ii) Stokes-Mueller calculus \cite{Azzam:16}, which describes polarization in terms of intensity-based parameters. Jones calculus assumes the light is coherent and therefore has difficulty representing depolarized light, which is common in experimental applications. For this reason, Stokes-Mueller calculus is more commonly used in experimental contexts.

The Stokes vector $S$ is a vector $4\times1$, which describes the state of polarized light in terms of four variables known as the Stokes parameters \cite{Azzam:16}. In term of canonical polarization states, it can be expressed as 
\begin{equation}
    S = \begin{bmatrix}
        I\\ Q\\ U\\ V
    \end{bmatrix}= \begin{bmatrix}
        \langle | E_x|^2 \rangle +  \langle | E_y|^2 \rangle\\  \langle | E_x|^2 \rangle-  \langle | E_y|^2 \rangle\\  \langle | E_{+45}|^2 \rangle -  \langle | E_{-45}|^2 \rangle\\ \langle | E_{RCP}|^2 \rangle - \langle | E_{LCP}|^2 \rangle\\
    \end{bmatrix}
    = \begin{bmatrix}
        I_H + I_V\\ I_H - I_V\\ I_{+45} - I_{-45}\\ I_{RCP} - I_{LCP}\\
    \end{bmatrix},
\end{equation}
where $I_H, I_V, I_{+45}, I_{-45}, I_{RCP}$ and $I_{LCP}$ are the measured intensities of the horizontal, vertical, $+45^{\circ}$, $-45^{\circ}$ linear, right circular and left circular polarizations respectively.

The interaction of light with a linear medium is represented by a matrix transformation of input polarized state $S_{i}$ as
\begin{equation}
    S_{o} = M S_{i}.
\end{equation}
Here $S_{o}$ is the output Stokes vector which describes the measured states of polarization and $M$ is $4 \times 4$ MM which describes the transfer function of the medium.

The determination of a MM requires at least 16 measurements, corresponding to all combinations of four distinct input and output polarization states, which are generated by a polarization state generator (PSG) and analyzed by a polarization state analyzer (PSA), respectively. We can gather intensity measurements in a matrix $I(x,y) = A M(x,y) W$, where $W$ and $A$ is a matrix of input and output Stokes vectors written in columns and in lines (matrices of PSG and PSA), and $x$ and $y$ indicate the position of a pixel. 

Note that, to improve accuracy, it is sometimes useful to increase the number of input and output Stokes vectors from $4$ to $N$. This changes the dimensions of $A$ from $4\times4$ to $N\times4$, $W$ from $4\times4$ to $4\times N$, and $I$ from $4\times4$ to $N\times N$. Only the dimension of the $M(x,y)$ remains unchanged and is always $4\times4$.

The MM per pixel $M(x,y)$ can be obtained using the Moore-Penrose inverse (also called a pseudoinverse) \cite{ghosh2011tissue}
\begin{equation}\label{eqq}
    M(x,y) = A^{\dagger}I(x,y)W^{\dagger}.
\end{equation}

However, since it is impossible to measure the MM elements directly, the calculation in Eq.~\eqref{eqq} does not yield equal errors for all MM elements. Many studies have focused on finding optimal sets of Stokes vectors \cite{smith2002optimization,he2015linear} and on controlling measurement errors \cite{ahmad2006error,tyo2002design} (see more references in review \cite{he2021polarisation}). However there is one crucial requirement that must always be satisfied: the set of Stokes vectors must be complete in order to obtain the MM from the transformation given in Eq.~\ref{eqq}.

\section{Backscattering single point illumination polarimetric imaging setups}

To achieve deep penetration into the sample, the illumination must be directed perpendicular to the interface, despite the fact that this interferes with the outgoing light. For this reason, an additional optical element is typically introduced to separate the illumination light from the collected backscattered signal. Backscattering scanning setups vary in design based on application requirements, desired spatial resolution, and the complexity of polarization analysis. A common approach is the point-scanning configuration, where a focused beam is raster-scanned across a sample, and the backscattered light is analyzed at each point to reconstruct the MM. These systems offer high spatial resolution and flexibility in polarization state generation and detection, but often require longer acquisition times due to their sequential scanning nature.

\begin{figure}[h]
    \centering
    \begin{subfigure}{0.32\textwidth}
    \centering
    \includegraphics[width=0.95\linewidth]{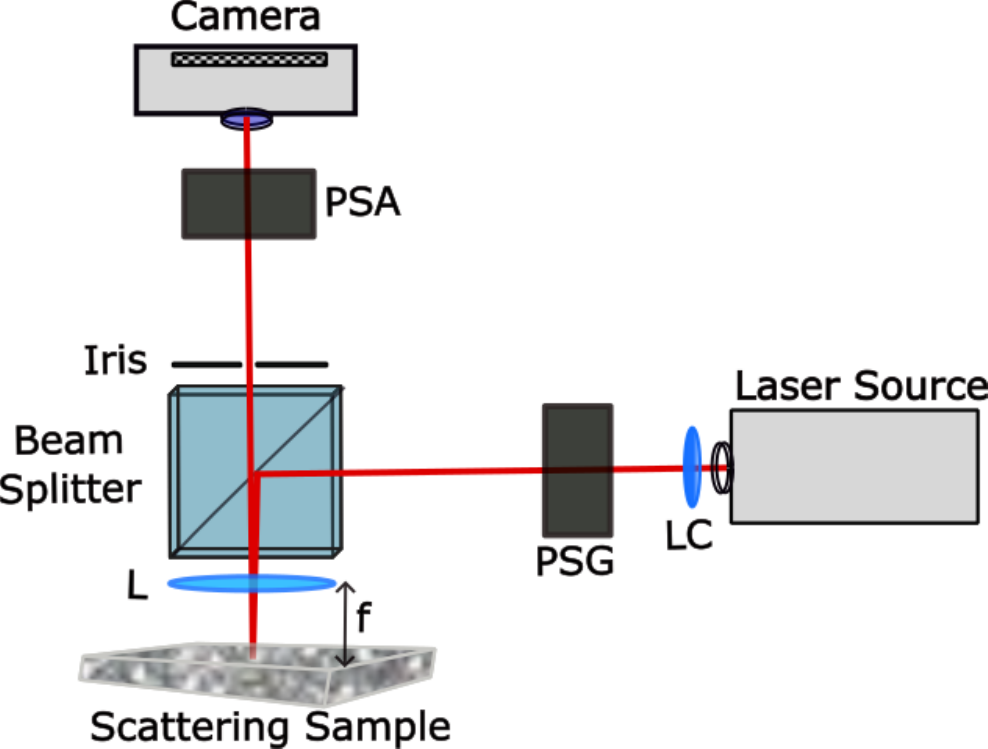}\\
    (a)
    \end{subfigure}
    \hfill
     \begin{subfigure}{0.32\textwidth}
    \centering
    \includegraphics[width=0.82\linewidth]{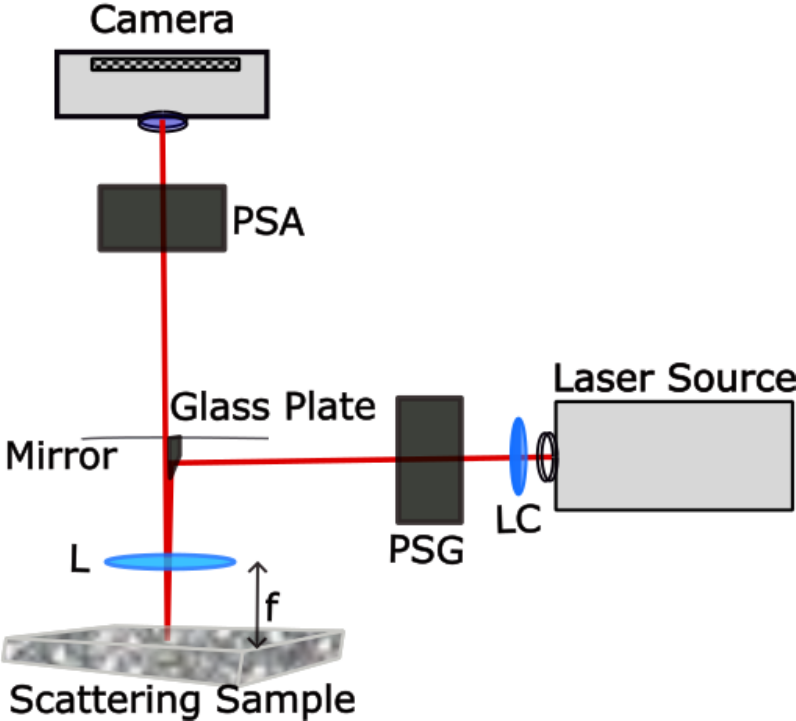}\\
    (b)
    \end{subfigure}
    \hfill
    \begin{subfigure}{0.32\textwidth}
    \centering
    \includegraphics[width=0.52\linewidth]{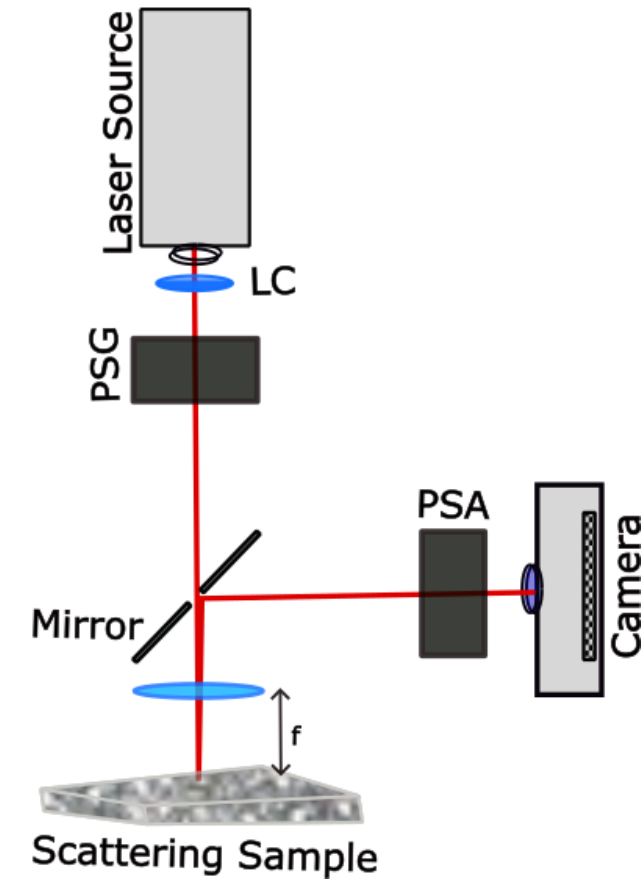}\\
    (c)
    \end{subfigure}
    \caption{Configurations for backscattering scanning imaging setups with an optical element, that separate incoming and outcoming light: (a) with a BS, (b) with a tiny mirror, (c) with a mirror with a central hole.}
    \label{fig:bsmu}
\end{figure}

At least three principal configurations are commonly used in backscattering scanning imaging setups (Fig.~\ref{fig:bsmu}). These include systems that employ a beamsplitter (BS) to separate incident and backscattered light (Fig.~\ref{fig:bsmu}(a)) \cite{Hornung:19, jiao2000depth,ni2004time}, a small mirror positioned to reflect the incident beam at a right angle (Fig.~\ref{fig:bsmu}(b)), and a planar mirror with a central hole (Fig.~\ref{fig:bsmu}(c)) \cite{hielscher1997diffuse, rakovic1999light} that allows the laser beam to pass through while reflecting the backscattered light. An alternative configuration using two mirrors has also been reported \cite{falconet2008analysis}, but it introduces a larger beam tilt, which is not optimal for collecting uniform backscattering.

The choice of optical configuration in backscattering scanning polarimetric imaging significantly affects the efficiency, accuracy, and practicality of the system. Among the commonly used approaches, each presents distinct advantages and limitations. The BS configuration is compact and relatively easy to align, but suffers from signal loss and potential polarization artifacts due to partial reflection and transmission. The tiny mirror and mirror-with-hole design can both avoid some of the BS’s losses, but they add mechanical complexity and require precise alignment to preserve integrity of the backscattered light, and they inherently fail to collect the most critical portion of the signal that travels directly back along the incident path. Therefore, in this study, we focus primarily on two configurations: (i) a setup using a BS  and (ii) a setup with a 45° mirror (Fig.~\ref{fig:setups}). 

\subsection{Scheme and description of setup with a BS}

\begin{figure}[h]
    \centering
    \begin{subfigure}{0.43\textwidth}
    \centering
    {\includegraphics[width=1\linewidth]{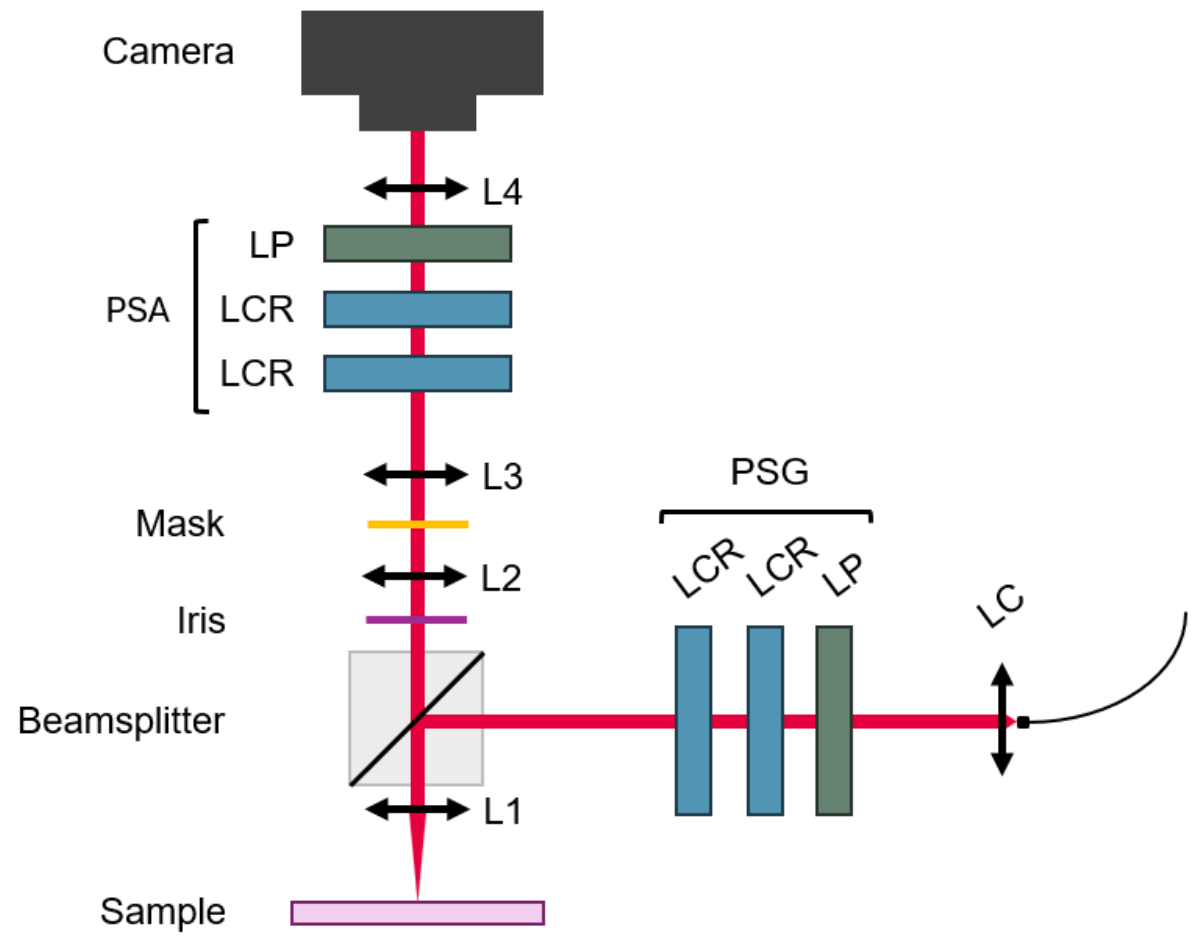}\\
    (a)}
    \end{subfigure}
    \hfill
     \begin{subfigure}{0.56\textwidth}
    \centering
    {\includegraphics[width=1.0\linewidth]{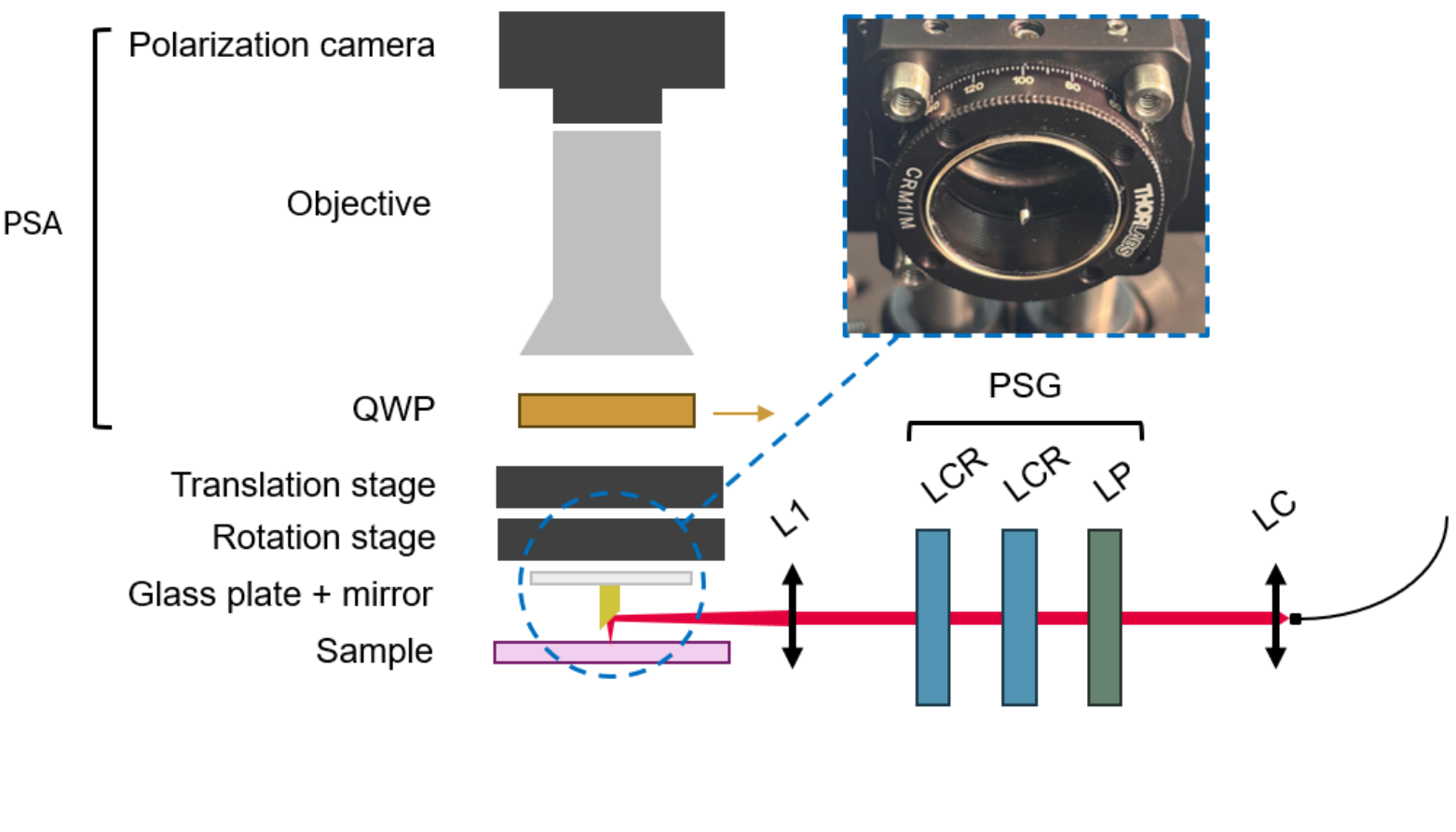}\\
    (b)}
    \end{subfigure}
    \caption{Polarimetric imaging setup in (a) BS-based configuration  and (b) mirror-based configuration.}
    \label{fig:setups}
\end{figure}

The standard polarimeter follows a common backscattering configuration with two perpendicular arms as shown in Fig.~\ref{fig:setups}(a). The horizontally oriented illumination arm includes the polarization state generator (PSG), while the vertical detection arm has a two-lens system with a polarization state analyzer (PSA).

The light source is a Picosecond Diode Laser LDH-IB-730-B (PicoQuant), coupled to a single-mode fiber, operating at $\lambda=727.6$ nm. The laser beam is collimated by a lens (LC) and directed onto the sample via a non-polarizing BS cube (50:50, polarization-maintaining within $0.5\%$ at a $\pi/4$ incidence angle). The PSG, acting on the collimated beam, generates definite polarization states using a linear polarizer (LP, Thorlabs, extinction ratio $1:100,000$) and two computer-controlled liquid crystal variable retarders (LCR, Meadowlark Optics). The polarized beam is then focused onto the sample using a lens (L1) with a 60 mm focal length. 

The beam radius at the sample is $36 \mu$m. The backscattered light from the sample passes through the BS and two-lens system comprising the lens L2 and an identical lens L3 (both with 30 mm of the focal distance). An iris at the back focal plane of L2 defines the numerical aperture (NA) of the setup, which is used to  minimize specular reflections for a beam.

The PSA is positioned between lens L4 (a 60 mm of focus distance) and L3. A CCD camera (ptGray Grasshopper, $16$-bit, $2448 \times 2048$ pixels) at the detection arm's end captures the backscattered light.

\subsection{Scheme and description of setup with a mirror}

The mirror-based polarimetric microscope consists of a light source, a PSG, a mirror system, and a PSA, as illustrated in Figure \ref{fig:setups}(b). The light source and PSG are identical to those in the BS setup. After passing through the PSG, the collimated laser beam is focused onto a small rod mirror (Edmund Optics, 2 mm length, 1 mm diameter, aluminum coated) using a lens (f = 75.0 mm). The back side of the rod mirror is fixed to a broadband anti-reflective coated glass plate (Edmund Optics, 25.0 mm diameter, 1.0 mm thickness). The mirror-glass plate assembly is mounted on a rotation mount and translation stage for precise alignment with the laser beam, ensuring the mirror is centered in the image as well as the captured backscattered light. This configuration blocks directly reflected light during image acquisition.

The PSA consists of a quarter-wave plate (QWP) (Thorlabs, 690 - 1200 nm), an objective lens (KOWA TV Zoom Lens Macro, focal length: 12.5 - 75 mm), and a polarization camera (FLIR Blackfly S BFS-U3-51S5P, 12-bit, $2448\times2048$ pixels). To measure circularly polarized light, a QWP is positioned between the mirror-glass plate assembly and the objective lens, requiring two sets of measurements, one with and one without the QWP to capture all six output polarization states. The objective lens focus is adjusted to obtain a sharp image of the mirror, resulting in a field of view of approximately $31.6 \times 26.4$ mm for the full image. 

To obtain the MM for the mirror-based setup a possible way is to combine intensity matrices from two sets $I_1$ and $I_2$ into a joint matrix $\tilde{I}$ and similarly combine the PSA Stokes vectors $A_1$ and $A_2$ into a joint matrix $\tilde{A}$. This doubles the number of lines, but allows to apply the same procedure as given in Eq.~\eqref{eqq} to define a MM. 

\subsection{Calibration of setups}

Both setups are calibrated using the maximum likelihood calibration method (see Appendix \ref{append}) described in \cite{Stefanov:25} with a polarizer and a QWP as reference elements. Six sets of measurements are carried out: (i) with a polarizer under a BS/mirror, (ii) with a polarizer under a BS/mirror and a QWP after PSG, (iii) with a polarizer under a BS/mirror and a QWP before PSA, (iv) with a QWP after PSG, (v) with a QWP before PSA and (vi) without reference elements.

For the setup with mirror, there is a need to perform a second pack of measurements with a QWP to define circular polarization elements.

\section{Experimental characterization of setups based on polystyrene suspension}

The primary goal of polarimetric backscattering setups is to investigate the scattering properties of highly diffusive biological tissues and to relate the measured parameters to their underlying microscopic structure.
To evaluate and compare the overall performance of the setups, it is useful to perform measurements on well-characterized and easily handled reference samples. In this work, quasi-monodisperse polystyrene particles diluted in deionized water are used as an etalon medium for polarimetric measurements. This choice is motivated by the fact that the particles are uncorrelated, optically inactive in water, and exhibit single-scattering properties that are well understood and extensively documented \cite{Hornung:19}. 
An additional key advantage of using polystyrene spheres is that their optical behavior can be accurately modeled using a variety of numerical approaches: (i) using Monte Carlo (MC) simulations \cite{Ramella-Roman:051, Ramella-Roman:052}, (ii) with the help of a numerical solution (for volume distributed function) of a vector radiative transfer equation \cite{Zege1991,tynes2001monte} or (iii) with the help of different analytical approximations \cite{cai2006analytical,liemert2017analytical,rakovic1999light,kim2011diffusion}, which aids in validating our experimental results. For this study, we use the MC simulation code described in \cite{gunhan2014monte1,gunhan2014monte2}.% with the corresponding parameters of interest are discussed ahead.

%\subsection{Model of scattering medium}

\subsection{Analysis of scattering based on de-rotated Mueller Matrix}

As a particular case of homogeneous media with randomly oriented particles, a suspension of polystyrene spheres satisfies symmetry requirements \cite{hulst1981light, hovenier1969symmetry}. This allows the angular dependence of the MM to be expressed in terms of the so-called de-rotated MM \cite{rakovic1999light} %$\tilde{M}(\rho)$:
\begin{equation}
{M}(\rho, \varphi) = J(\varphi) \, \widetilde{M}(\rho) \, J(\varphi),
\label{spatial_mueller}
\end{equation}
where $\rho$ and $\varphi$ are polar coordinates in plane of CCD, and the rotation matrix $J(\varphi)$ is given by
\begin{equation}
    J(\varphi) = \begin{bmatrix}
        1 & 0 & 0 & 0\\
        0 & \cos{(\varphi)} & -\sin{(\varphi)} & 0\\
        0 & \sin{(\varphi)} & \cos{(\varphi)} & 0\\
        0 & 0 & 0 & 1
    \end{bmatrix}.
\end{equation}

The rotation matrix $J(\varphi)$ accounts for the change in polarization orientation observed from different reference frames. While polarization states are defined in a fixed global frame, the scattering process, assumed to be rotationally invariant, is described in a local frame rotated by the azimuthal angle $\varphi$ relative to the global frame. The first rotation matrix transforms the polarization state from the global to the local frame, and the second transforms it back to the global frame. In backscattering geometry, where incident and scattered beams propagate in opposite directions, the rotation angles on both sides of the intrinsic matrix have the same sign.

The de-rotated MM can be obtained from the spatial MM by inverting Eq.~\eqref{spatial_mueller}:
\begin{equation}
\widetilde{{M}}(\rho) = J(-\varphi) \, {{M}}(\rho, \varphi) \, J(-\varphi).
\label{derotated_mueller}
\end{equation}

\begin{figure}[h]
    \centering
    \begin{subfigure}{0.32\textwidth}
        \centering
        \includegraphics[width=0.965\linewidth]{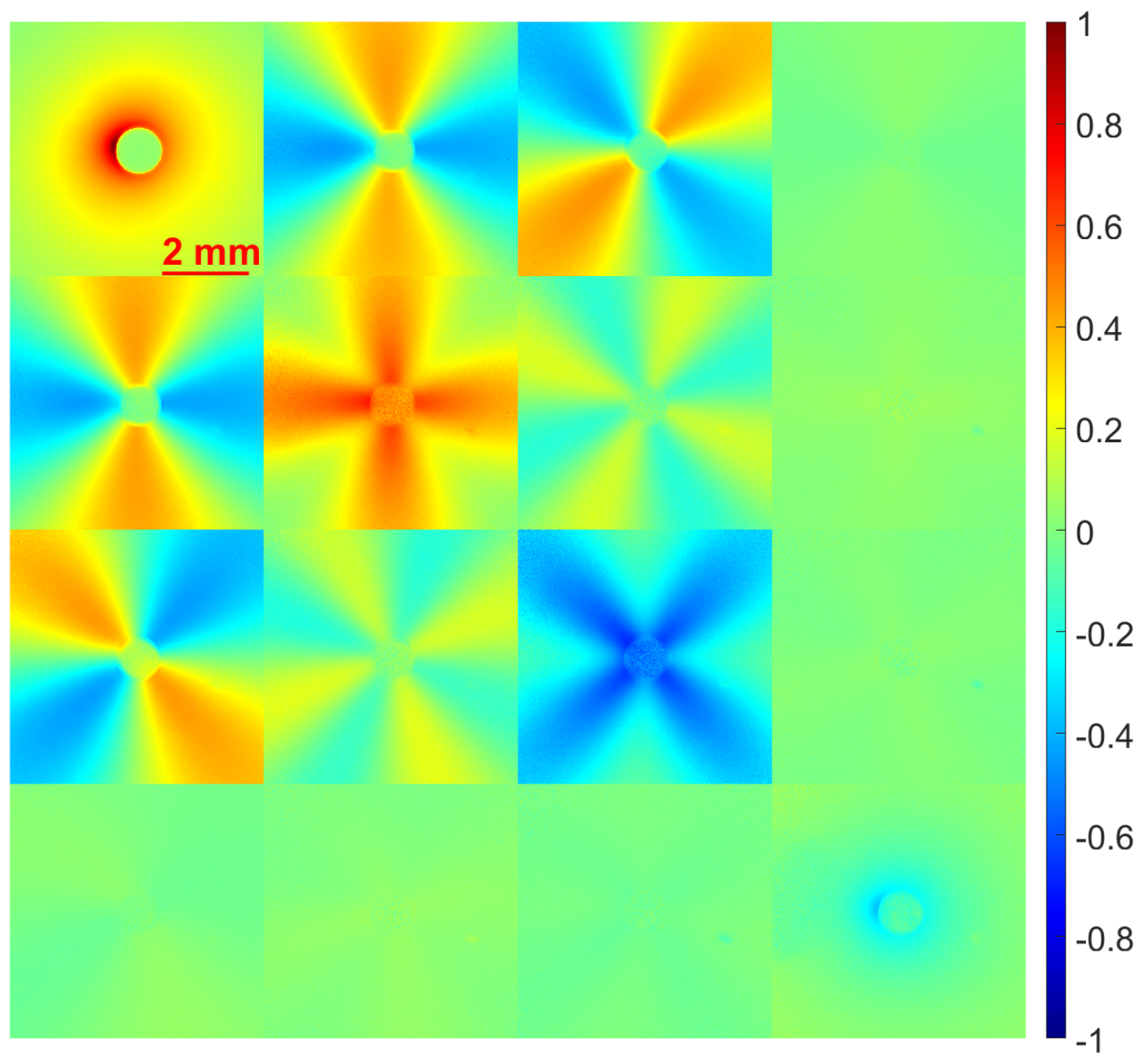}
        \\(a)    
        \end{subfigure}
    \hfill
    \begin{subfigure}{0.32\textwidth}
        \centering
        \includegraphics[width=1\linewidth]{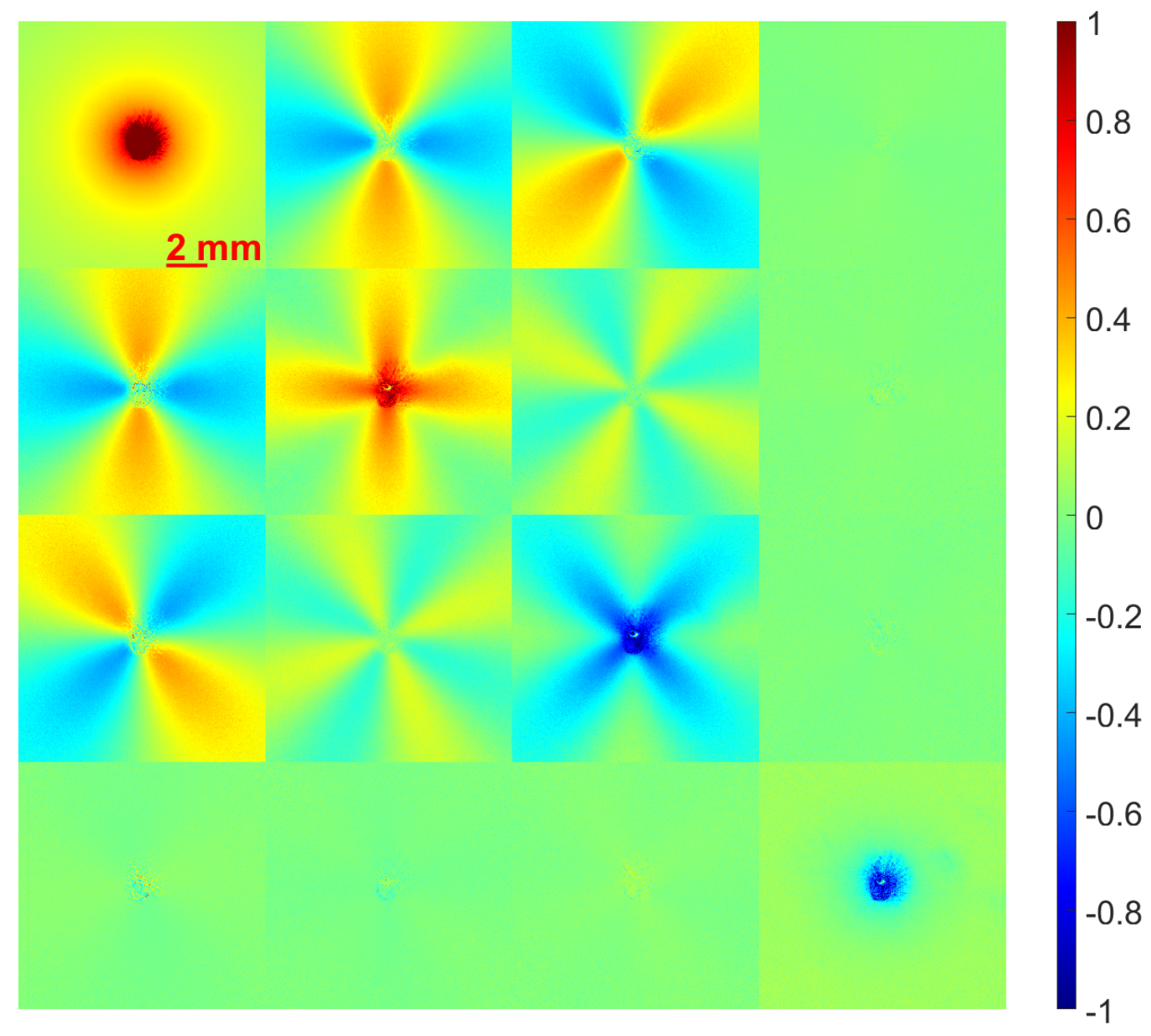}
        \\(c)
    \end{subfigure}
    \hfill
    \centering
    \begin{subfigure}{0.32\textwidth}
        \centering
        \includegraphics[width=0.96\linewidth]{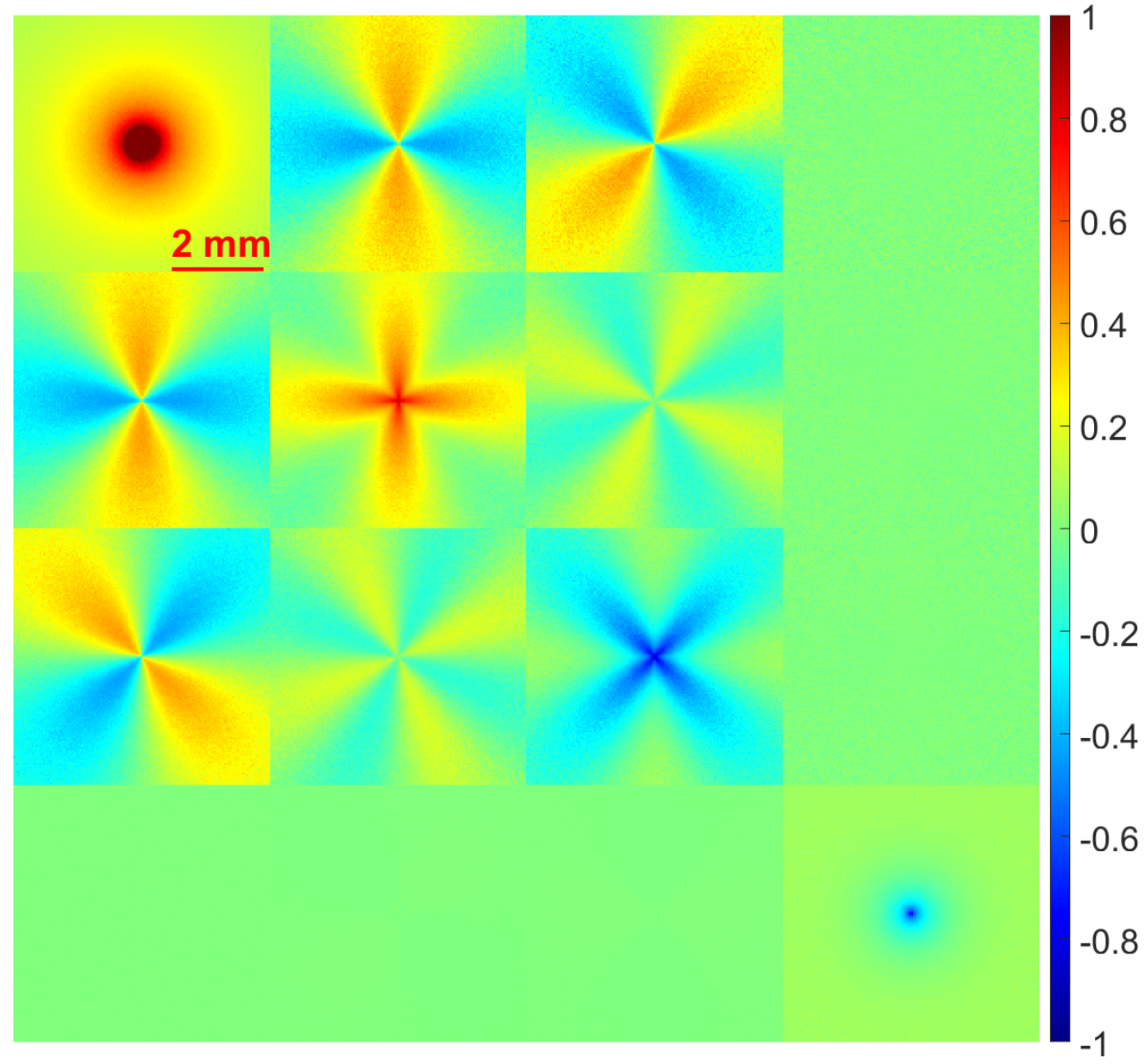}
        \\(e)
    \end{subfigure}    
    \hfill
    \begin{subfigure}{0.32\textwidth}
        \centering
        \includegraphics[width=0.965\linewidth]{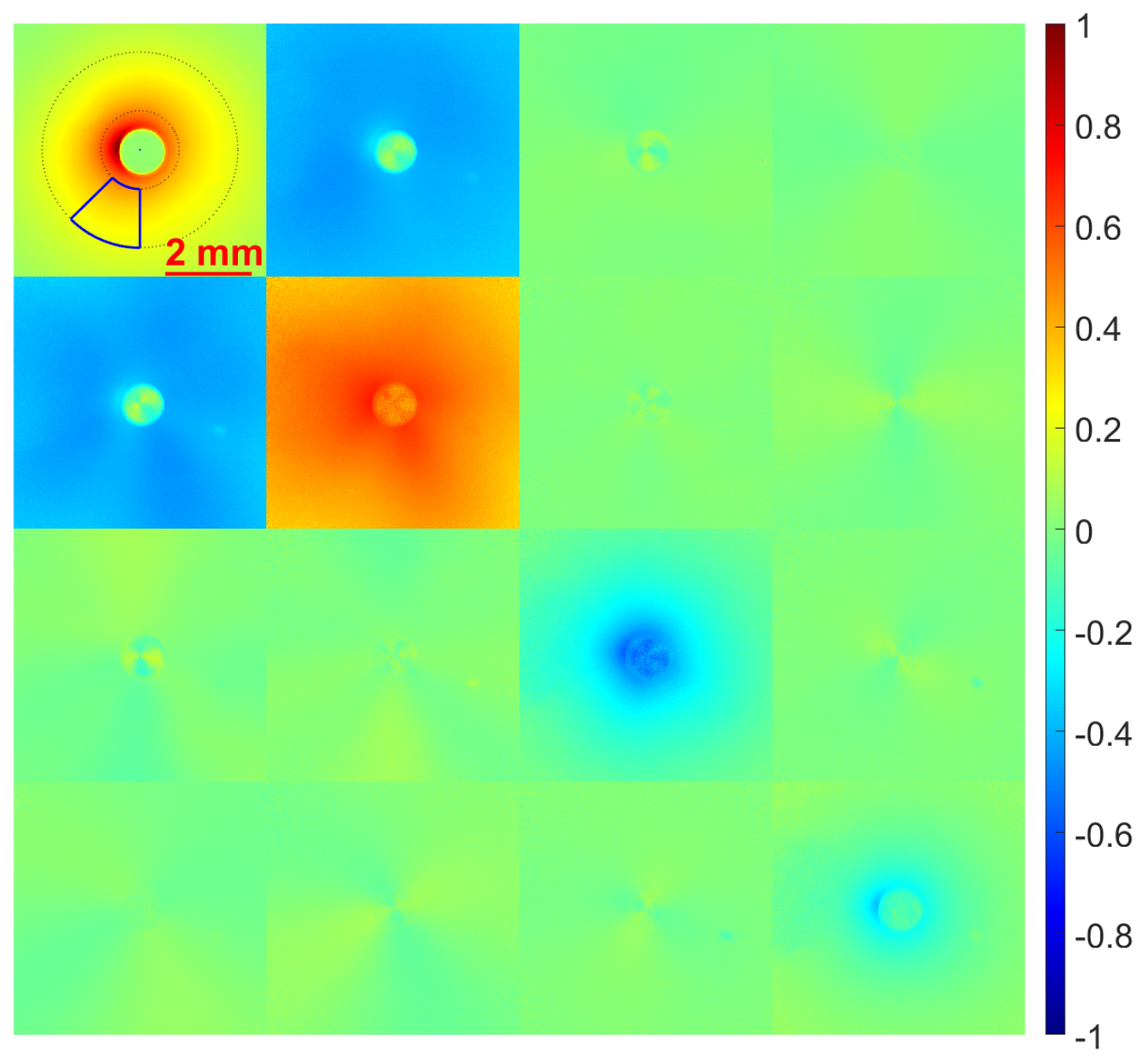}
        \\(b)        
    \end{subfigure}
        \hfill
    \centering
    \begin{subfigure}{0.32\textwidth}
        \centering
        \includegraphics[width=1\linewidth]{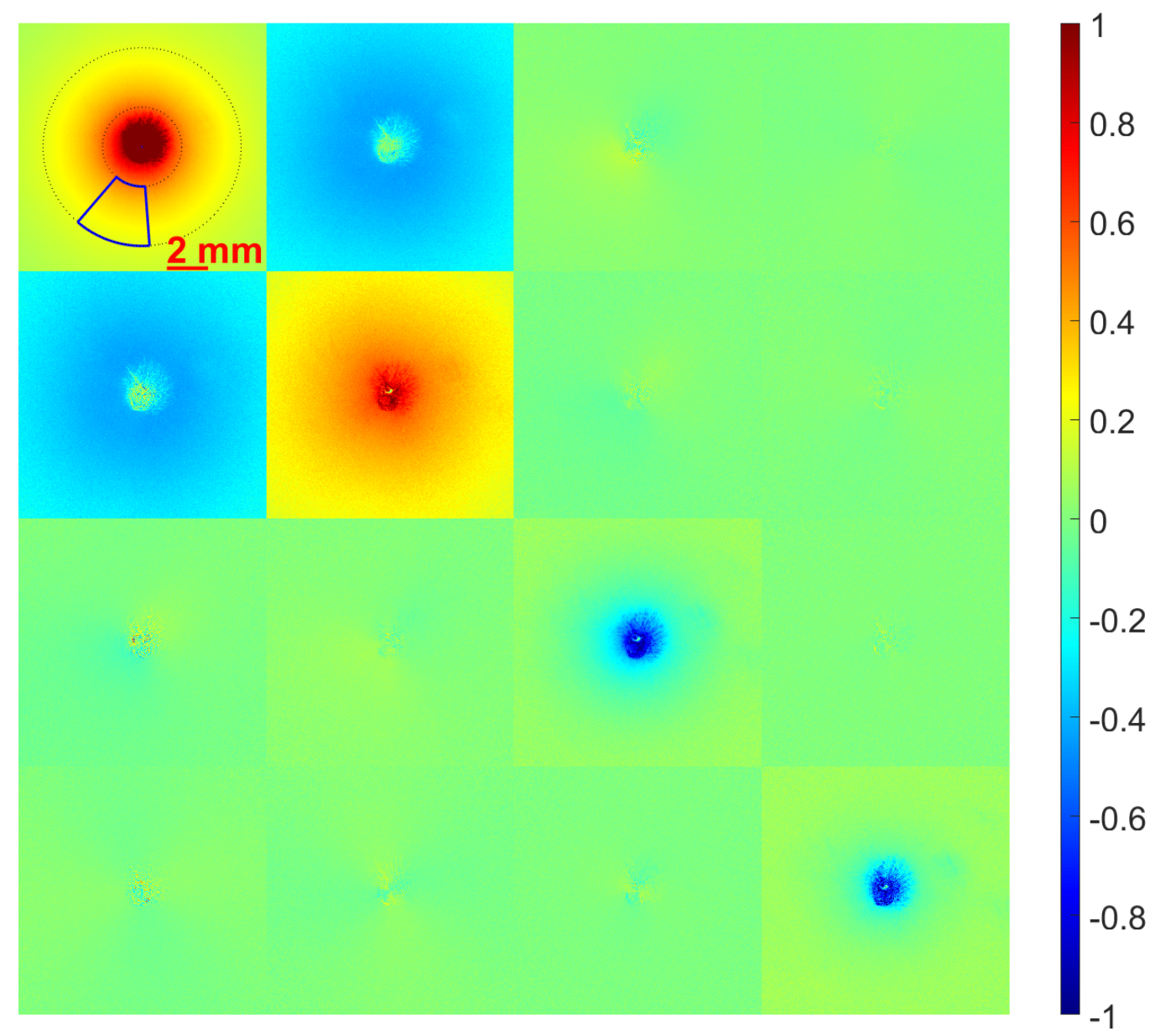}
        \\(d)
        
    \end{subfigure}    
    \hfill
    \begin{subfigure}{0.32\textwidth}
        \centering
        \includegraphics[width=0.96\linewidth]{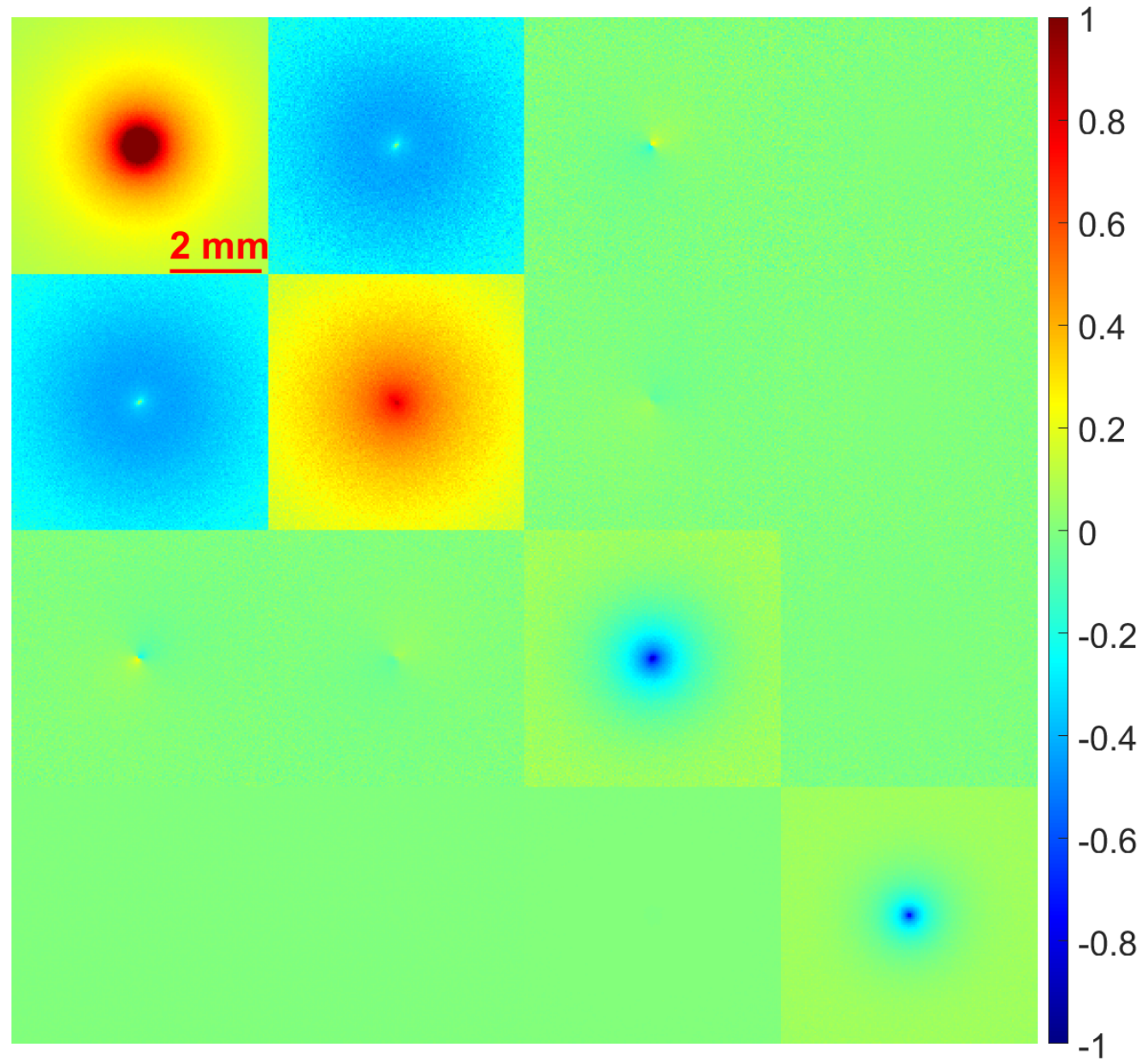}
        \\(f)
    \end{subfigure}
    \caption{
    (a) and (b) are MM and de-rotated MM respectively as measured by BS setup for $R=320\,\ \mathrm{nm}$ and $\mu_s=0.461\,\ \mathrm{mm}^{-1}$; (c) and (d) are MM and de-rotated MM respectively as measured by mirror setup for $R=320\,\ \mathrm{nm}$ and $\mu_s=0.461\,\ \mathrm{mm}^{-1}$; (e) and (f) are MM and de-rotated MM respectively calculated by MC simulation for $x=1.8$ and $\mu_s=1\,\ \mathrm{mm}^{-1}$. The element $\widetilde{M}_{11}$ of experimental de-rotated MM is marked with region of interest, which we use in all MM elements to avoid measurement noise and internal reflections in (b) and (d). %The marked region is shown only on the first component $\widetilde{M}_{11}$ of the de-rotated MM, but we used the same region on each element for our calculations. 
    All elements are normalized by $M_{11}$, whereas the $M_{11}$ element itself is normalized by the maximum value $M_{11}(\rho_0)$, covering the intensity scale in range $\rho\ge\rho_0$. Values $\rho_0$ are taken at $\rho_0 = 0.5\,\mathrm{mm}$ for (a), $\rho_0 = 1.0\,\mathrm{mm}$ for (c), and by an averaged value at $\rho_0 = 0.4\,\mathrm{mm}$ for (e). }
    \label{fig:MM}
\end{figure}

The de-rotated MM depends solely on the radial distance $\rho$ from the point of illumination, highlighting the rotational symmetry of the colloidal suspensions (an example is shown in Fig.~\ref{fig:MM}). This structure is formed by the inherent reciprocity and mirror symmetry of the scattering behavior exhibited by the individual polystyrene spheres, and has the form of Eq.~\eqref{MM_sim} \cite{rakovic1999light}
\begin{equation}\label{MM_sim}
    \widetilde{M}(\rho) = \begin{bmatrix}
        a_1(\rho) & b(\rho) & 0 & 0\\
        b(\rho) & a_2(\rho) & 0 & 0\\
        0 & 0 & d_1(\rho) & c(\rho)\\
        0 & 0 & -c(\rho) & d_2(\rho)
    \end{bmatrix}.
\end{equation}
Thus, the de-rotated MM contains six independent radial functions (see Fig.~\ref{fig:MM} (b,d,f)).

The scattering model assumes randomly distributed spherical scatterers governed by Mie theory. It involves four parameters: $\mu_s$ and $\mu_a$, which define the scattering and absorption coefficients (related to the concentration of the suspension and its absorption characteristics); the relative refractive index $n_r = n_s / n_m$, which is the ratio between the refractive indices of the scattering spheres ($n_s$) and the surrounding medium ($n_m$); and the size parameter $x = k r$, which is the product of a wave number of the illuminating light and the radius of the sphere.

In the diffusive approximation the model of scattering depends on only one parameter \cite{Martelli2009LightPT} $\mu_s' = \mu_s(1 - g)$, where $g$ is the anisotropy factor representing the average cosine of the scattering angle (the anisotropy factor is uniquely determined by the parameters $x$ and $n_r$, and for a fixed $n_r$, $g$ can be calculated from $x$). Due to this, we must first establish whether it is fundamentally possible to recover the full set of unknown parameters. To address this, we use the formalism of Fisher information (FI) \cite{frieden2004science,schervish2012theory}.

\subsection{Estimation of errors based on Fisher information framework}\label{sect:FI}

By definition, the FI measures the overall sensitivity of the functional relationship $f(x|\nu)$ to changes in parameter $\nu$, by weighting the sensitivity in each potential outcome $x$ with respect to the probability defined by $p_\nu(x) = f(x|\nu)$ \cite{schervish2012theory}. It can be expressed in the form
 \begin{equation}
 \mathfrak{F}_{nm}=\int\limits_{M}^{} dx\left(\frac{\partial}{\partial \nu_n} \log f(x|\nu) \right)\left(\frac{\partial}{\partial \nu_m} \log f(x|\nu) \right)p_\nu(x).
\end{equation}

For a continuous set of 4 variables $\nu=\{\mu_a,\mu_s\,x,n_r\}$ with each particular measurement, we can assume that the intensity is normally distributed (with variance $\sigma^2$) and independent of other measurements. Additionally, we can assume that the variance is identical, regardless of the intensity value. Under these assumptions, the FI  \cite{schervish2012theory} simplifies to
\begin{equation}\label{fisher}
    \mathfrak{F}_{nm}=\frac{1}{\sigma^2}\sum_{i,j,\rho}\frac{\partial}{\partial \nu_n}\left(\widetilde{M}_{ij}(\rho)\right)\frac{\partial}{\partial \nu_m}\left(  \widetilde{M}_{ij}(\rho)\right).
\end{equation}
The diagonal elements of the inverse FI represent a lower bound on the variance of the corresponding variables, known as the Cramér-Rao bound (CRB) \cite{frieden2004science}
\begin{equation}\label{cramer-rao}
    \Delta^2 \mu_n\ge(\mathfrak{F})^{-1}{}_{nn}.
\end{equation}

Thus, the existence of the inverse FI indicates the possibility of estimating $\nu$, while the magnitude of its diagonal elements reflects the relative error associated with the parameters.

When the de-rotated MM is expressed as an analytical or numerical function of the parameter $\nu$, the associated estimation errors can be evaluated directly using Eq.~\eqref{cramer-rao}. However, the most common approach for computing the MM is through MC simulations, which yield a discrete and noisy representation of the MM. Since derivative calculations are required, this introduces significant challenges, particularly in interpolation and subsequent data processing. To substantiate this challenge, we consider MM of MC simulation in a couple with an approximated theoretical model of the de-rotated MM $\widetilde{M}(\rho)$, based on observed patterns arising from double scattering of light by spheres \cite{rakovic1998theoretical}

\begin{equation}\label{MM_approx}
    \widetilde{M}^{(appr)}(\rho)=\frac{\mu_s^2}{2(\mu_s+\mu_a)\rho}\int\limits_0^{\pi/2}d\theta e^{-(\mu_s+\mu_a)\rho \cot{\theta/2}}\left(M(\theta)M(\pi-\theta)+M(\pi-\theta)M(\theta)\right),
\end{equation}
where $M(\theta)$ is the well-known Mie scattering MM\cite{hulst1981light}. 

\begin{figure}[h]
\begin{minipage}[b]{0.49\linewidth}
  \centering
  \includegraphics[width=1\columnwidth]{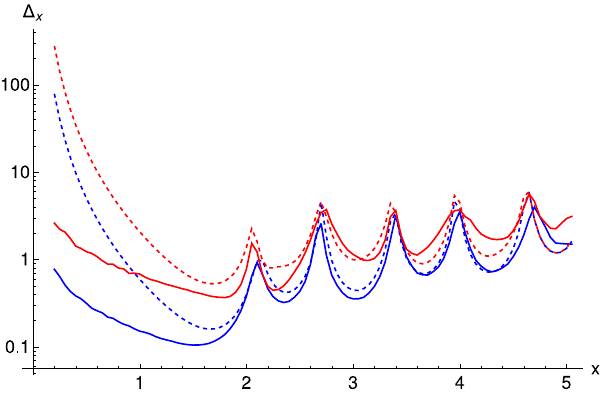}  \\
  (a)
\end{minipage}
\begin{minipage}[b]{0.49\linewidth}
  \centering
   \includegraphics[width=1\columnwidth]{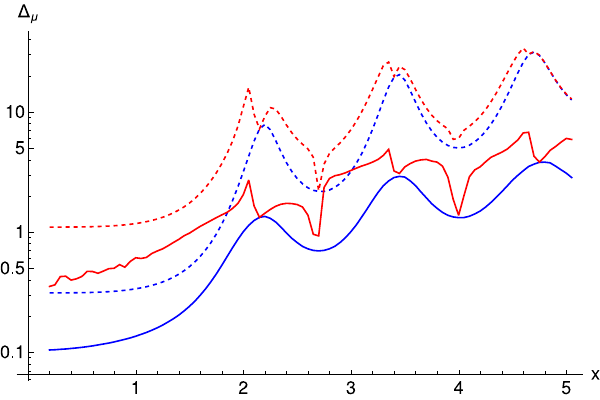} \\
  (b)
\end{minipage}
\caption{
Standard deviations $\Delta_x$ (a) and $\Delta_{\mu}$ (b), calculated from the CRB (Eq.~\eqref{cramer-rao}) for the de-rotated MM obtained via a simulation approach (solid lines) and for the approximate theoretical model of the de-rotated MM (Eq.~\eqref{MM_approx}) (dashed lines), when one variable is unknown (blue curves) or both parameters are unknown (red curves). The plots show the dependence on $x$ with $\mu_s = 1\,\mathrm{mm}^{-1}$. The simulations are performed using $10^8$ photon realizations for $x$ from $0.2$ to $5.0$ with step $0.05$, $\mu_s=1mm^{-1}$, counted on CCD array $12mm\times12mm$ with the size of pixel equals to $0.04mm$. The range of summation for $\rho$ is $[0.08mm,5.96mm]$ (the step is $0.04mm$) with normalizing of the both MM by a constant value $\tilde{M}_{11}$ of correspondent MM in $\rho=0.08mm$ for $x=2$. }
\label{fig:errors}
\end{figure}

To simplify the calculation of the FI based on MC simulations, a few assumptions are made. First, the number of variables is reduced by assuming that $n_r$ is known and $\mu_a=0$ (at the experimental wavelength used, the absorption coefficient of the medium is small $\mu_a\approx10^{-3}$ and can be neglected). Additionally, a constraint is assumed between the polar coordinates $\rho$ and the coefficients $\mu_s$, because of a similarity principle (or a scale property of scattering) \cite{Zege1991,Martelli2009LightPT}
\begin{equation}
    \tilde{M}^{(sim)}(\rho)=\mu_s^2 \tilde{M}\left((\mu_s+\mu_a)\rho\right)\approx \mu_s^2 \tilde{M}\left(\mu_s\rho\right).
\end{equation}
This trick allows the reuse of a single simulation for different values of parameter $\mu_s$.

The standard deviations for $x$ and $\mu_s$ are shown in Fig.~\ref{fig:errors}. Solid lines correspond to the simulation results, while dashed lines represent the approximate ones. When only one parameter is known (blue), the errors are consistently smaller compared to the case where the both parameters are unknown (red). Since FI is additive, the approximate model generally yields larger errors than the exact model, as it includes only leading-order scattering contributions. However, a normalization to a constant value is applied so that both cases share the same scale, which explains the intersection between the solid and dashed curves. %As expected, the case with two unknown parameters (red lines) yields a larger error than the case with one unknown parameter (blue lines).

The main conclusion from Fig.~\ref{fig:errors} is that the MM contains sufficient information to estimate the unknown parameters with finite accuracy. This dependence is influenced by the standard deviation of a setup noise $\sigma$, which decreases with the number of measurements $N$ as $1/\sqrt{N}$.

\subsection{Experimental estimation of the polarization setups accuracy}

It is therefore possible to estimate unknown parameters of the polystyrene suspension from the de-rotated MM, and we can try to do it experimentally. The maximum likelihood method is applied under the assumption of Gaussian-distributed errors in the experimental data. The expression for a maximization problem takes the form (see Appendix \ref{append} for details)
\begin{equation}\label{ML_method}
        \mathfrak{l}(\zeta)=\frac{\left(\sum\limits_{i=1}^{6} \sum\limits_{\rho=\rho_1}^{\rho_2}  \tilde{M}_i^{(sim)}(\rho) \tilde{I}_i(\rho)^T \right)^2}{\sum\limits_{i=1}^{6}\sum\limits_{\rho=\rho_1}^{\rho_2} (\tilde{M}_i^{(sim)}(\rho))^2 },
\end{equation}
where $\tilde{I}(\rho)$ is the de-rotated MM extracted from the experimental data, and the index $i$ runs over all non-zero elements of the de-rotated MM: $M_{11}, (M_{12}+M_{21})/2, M_{22}, M_{33}, (M_{34}-M_{43})/2, M_{44}$, as follows from Eq.~\eqref{MM_sim}. As mentioned above, integration over the polar coordinate $\rho$ should be replaced by a summation with a step size equal to the pixel size, due to the discrete nature of the experimental data.

It is worth noting that the simulations assume point-like illumination, whereas in the experiments the beam has a finite spatial profile, typically a TEM$_{00}$ mode. This effect can be modeled as a superposition of ray illuminations at different spatial positions with the Gaussian weights. When using a reference dataset stretched by the factor $\mu_s$, the Gaussian beam width $\omega$ must be rescaled accordingly $\omega = \omega_0/\mu_s$, as %from the following:
\begin{multline} 
\widetilde{M}_G^{(sim)}(\mu_s\rho,\omega)\sim\int d\bm{r} 
e^{-2 |\bm{r}|^2/\omega_0^2}\widetilde{M}^{(sim)}\left(|\mu_s\bm{\rho}+\bm{r}|\right) \sim \\
\sim \int d\bm{r} e^{-2 \mu_s^2|\bm{r}-\bm{\rho}|^2/\omega_0^2}\widetilde{M}^{(sim)}\left(\mu_s|\bm{r}|\right). 
\end{multline}

Nevertheless, in the described setups the beam waist is $\omega_0<0.1\,\mathrm{mm}$ , which allows the effect of the finite beam size to be neglected.

Another important comment about Eq.~\eqref{ML_method} is that we apply either a mask to block the reflection of the laser beam from the surface or observe a natural shadow caused by the mirror (or by the hole, as in setups such as Fig.~\ref{fig:bsmu}(c)). Consequently, the summation over $\rho$ in Eq.~\eqref{fisher} begins at a finite value $\rho_1$, rather than at $\rho = 0$. Since the observation region is physically bounded, the summation ends at $\rho_2$.

The experiment is performed on a polystyrene suspension with a scatterer diameter of $d = 0.32\,\mu\mathrm{m}$ (size parameter $x_0 = 1.8369$), carried out for three values of ${\mu_s}_0$: $0.461$, $0.817$, and $1.037$ $mm^{-1}$. The working range for the BS-based setup is taken as $[240\,\mathrm{px},\,600\,\mathrm{px}]$, with a pixel size of $0.0037\,\mathrm{mm}$, and 
for the mirror-based setup is as $[80\,\mathrm{px},\,200\,\mathrm{px}]$, with a pixel size of $0.0241\,\mathrm{mm}$. Due to reflection and transmission artifacts of the laser beam on internal optical elements of the setups, we use only a sector with an angular size of $\pi/4$, which is fixed for all measurements, as shown in Fig.~\ref{fig:MM}. This selection is based on empirical inspection and is fixed for all datasets acquired with a given setup.  Although using the full area yields reasonably good results (e.g. $x = 1.7548$ with $\mu_s = \{0.488$, $0.804$, $1.039\}\,\mathrm{mm}^{-1}$ for the BS-based setup and $x = 1.7651$ with $\mu_s = \{0.444$, $0.806$, $1.030\}\,\mathrm{mm}^{-1}$ for the mirror-based setup), this selection procedure can be applied to avoid internal reflections in the measurements to increase accuracy of calculation.

The reference database is based on the same simulations used for the calculation of the FI (see the caption of Fig.~\ref{fig:errors}). It consists of $10^8$ photon runs with $x$ in the range from $0.2$ to $5.05$ with a step of $0.05$, $\mu_s = 1\,\mathrm{mm^{-1}}$, and recorded on a CCD of size $12\,\mathrm{mm} \times 12\,\mathrm{mm}$ with a pixel size of $0.04\,\mathrm{mm}$.

\begin{figure}[h]
    \centering
        \begin{subfigure}{0.32\textwidth}
        \centering
        \includegraphics[width=1\linewidth]{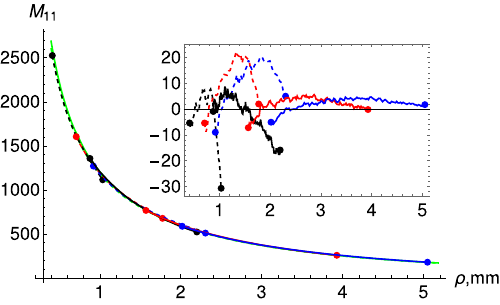}\\
        (a)
        \end{subfigure}
    \hfill            \begin{subfigure}{0.32\textwidth}
        \centering
        \includegraphics[width=1\linewidth]{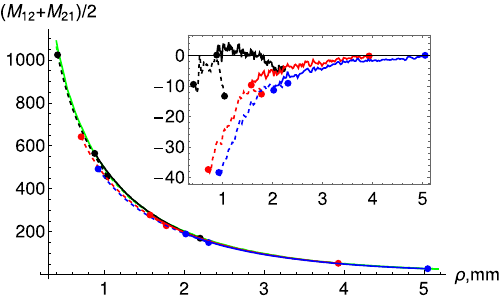}
        \\(b)
        \end{subfigure}
    \hfill
        \begin{subfigure}{0.32\textwidth}
        \centering
        \includegraphics[width=1\linewidth]{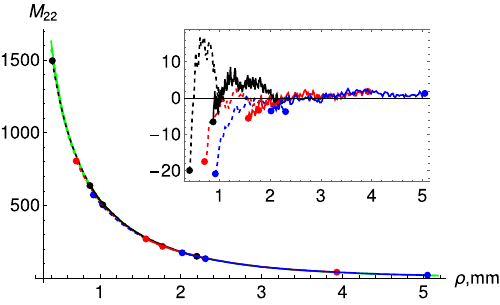}
        \\(c)
        \end{subfigure}
    \hfill
    \centering
        \begin{subfigure}{0.32\textwidth}
        \centering
        \includegraphics[width=1\linewidth]{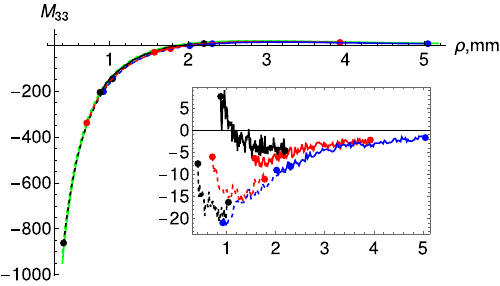}\\
        (d)
        \end{subfigure}
    \hfill    
        \begin{subfigure}{0.32\textwidth}
        \centering
        \includegraphics[width=1\linewidth]{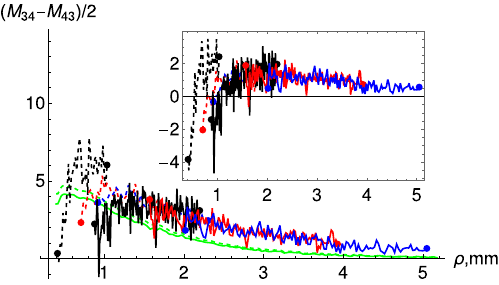}
        \\(e)
        \end{subfigure}
    \hfill
        \begin{subfigure}{0.32\textwidth}
        \centering
        \includegraphics[width=1\linewidth]{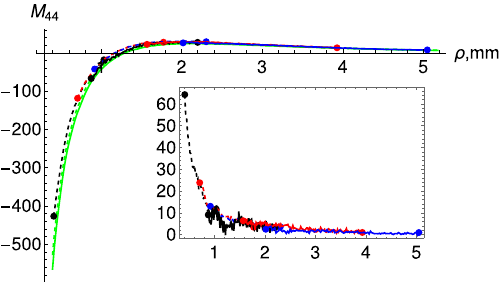}
        \\(f)
        \end{subfigure}
    \hfill
    \caption{Experimental de-rotated MM data with the polar coordinate rescaled by the scattering coefficient $\mu_s$ for the polystyrene suspension ($x_0 = 1.837$) (averaged over 9 measurements); $\mu_s = \{0.457$, $0.816$, $1.048\}\,\mathrm{mm}^{-1}$ for the mirror-based setup (blue, red, and black solid lines) and $\mu_s = \{0.467$, $0.800$, $1.038\}\,\mathrm{mm}^{-1}$ for the BS-based setup (blue, red, and black dashed lines). Green solid and dashed lines indicate simulations computed with the corresponding fitted values of $x = 1.7734$ (solid) and $x = 1.7836$ (dashed) with $\mu_s = 1$. Insets show the difference between experimental segments and the fitted line. }
    \label{fig:fitting}
\end{figure}

The reconstruction results are presented in Tab.~\ref{tab_data}. Both setups exhibit statistical errors below $0.2\%$ in the estimation of $x$ and $\mu_s$. However, a small systematic error is observed in the determination of the scatterer size, amounting to approximately $12\,\mathrm{nm}$ in diameter. Since this error is consistent across both setups (see Fig.~\ref{fig:fitting}), it is likely related to sample uncertainty, as the manufacturer (Magsphere: Uniform Polystyrene Latex $0.32\,\mu\mathrm{m}$) specifies a diameter tolerance of approximately $10\%$.

\begin{table}[htb]\label{tab_data}
\centering\caption{Restored parameters $x$ and $\mu_s$ of a polystyrene suspension. Set for each $\mu_s$ consists of 9 measurements, a standard deviation is $\Delta x<0.0006$, $\Delta \mu_s<0.003\,\mathrm{mm}^{-1}$ for both setups.  }
\begin{tabular}{|c|c|c|c|c|c|c|}
\hline
$d$, nm & $x_0$& ${\mu_s}_0$, mm$^{-1}$ & $x^{(BS)}$ & $x^{(mir)}$& $\mu_s^{(BS)}$, mm$^{-1}$ & $\mu_s^{(mir)}$, mm$^{-1}$ \\ 
\hline

320 & 1.8369  & 0.461  &1.7935   &1.7772 &0.457 &0.455\\ 
    &         & 0.817  &1.7871   &1.7760 &0.796 &0.816\\ 
    &         & 1.037  &1.7790   &1.7689 &1.042 &1.045\\ 
\hline
\end{tabular}\label{tab_data}
\end{table}

Due to the similarity principle discussed above for polystyrene suspensions, stretching the radial coordinate of the experimental datasets by the factor $\mu_s$ allows %the datasets to be fitted by 
us to fit the data by
a single curve (Fig.~\ref{fig:fitting}). This enables to join the datasets obtained for different values of $\mu_s$.
The resulting parameters are $x = 1.7734$ and $\mu_s = \{0.457, 0.816, 1.048\}\,\mathrm{mm}^{-1}$ for the mirror-based setup; scaling the experimental data as $\rho \rightarrow \mu_s \rho$ yields the corresponding blue, red, and black solid lines, and for the BS-based setup, $x = 1.7836$ with $\mu_s = \{0.467, 0.800, 1.038\}\,\mathrm{mm}^{-1}$, producing the corresponding blue, red, and black dashed lines. The green lines correspond to simulations with the same $x = 1.7734$ (solid), $x = 1.7836$ (dashed) and $\mu_s = 1\,\mathrm{mm}^{-1}$.

Fig.~\ref{fig:fitting} (e) shows the averaged element $\widetilde{M}_{34}$ of the de-rotated MM. Its magnitude is approximately two orders of magnitude smaller than the dominant elements and close to the noise level, which explains its instability.

\section{Discussion and conclusions}

We have presented here two configurations of backscattering single point illumination polarimetric setups. Thanks to accurate polarization calibration and modeling through MC simulation, we were able to retrieve the scattering parameters of the reference media from the measurements. The discrepancy between the simulated and measured values can be attributed to imperfect knowledge of the scatterers but also to imperfection of the setups, the main cause of errors being unwanted reflections. It has to be noted that the large dynamical range between the illumination and the backscattered light makes any backscattering measurement very sensitive to imperfection in the optical setup.

The characterization of the media was done by fitting the experimental polarimetric data with a scattering model. We apply this framework to polystyrene suspensions, whose scattering properties can be modeled using the Mie theory. As seen in Fig.~\ref{fig:fitting}, the agreement between the experimental and fitted data is very good for both setups.

In general, the scattering medium will be characterized by more parameters. For instance, scattering by oriented anisotropic scatterers would be parametrized by the orientation angle and degree of alignment, in addition to the scatterer size and the concentrations. The database needed for the fitting function will then be larger, but can still be built upon MC simulations. The subsequent application of the FI framework (Sect.~\ref{sect:FI}) to the database makes it possible to estimate parameter uncertainties and characterize the effectiveness of the measurements for the considered model even before the experiment. For example, Fig.~\ref{fig:errors}(a) shows a minimum in the measurement error of the scatterer size at approximately $x \approx 1.6$. This indicates that an appropriate choice of the wavelength can provide higher accuracy for a fixed scatterer size.

\begin{figure}[h]
    \centering
    \includegraphics[width=1\linewidth]{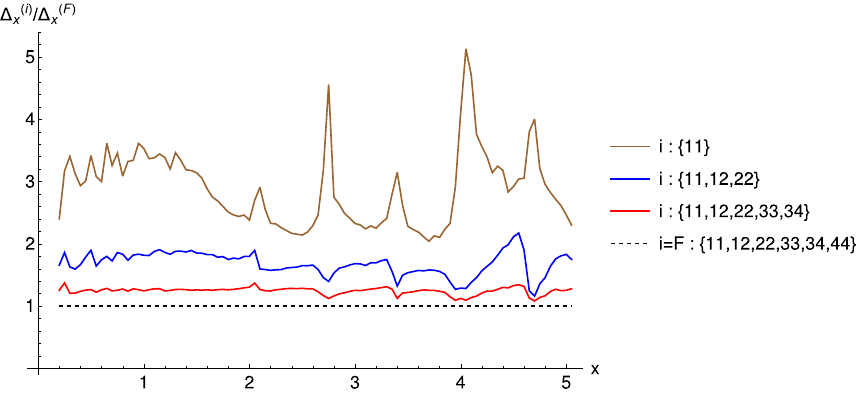}
    \caption{Ratios of the standard deviations obtained using different numbers of de-rotated MM elements relative to the full set, assuming that both parameters $x$ and $\mu_s$ are unknown. All values are calculated from the CRB (Eq.~\eqref{cramer-rao}) for the de-rotated MM obtained in the simulation approach, as a function of $x$ with $\mu_s = 1\,\mathrm{mm}^{-1}$. The MC simulation consists of $10^8$ photon runs for $x$ ranging from $0.2$ to $5.05$ with a step of $0.05$, with $\mu_s = 1\,\mathrm{mm}^{-1}$. Photons are counted on a CCD of size $12\,\mathrm{mm} \times 12\,\mathrm{mm}$ with a pixel size of $0.04\,\mathrm{mm}$. The summation range for $\rho$ is $[0.08\,\mathrm{mm},\,5.96\,\mathrm{mm}]$ with a step of $0.04\,\mathrm{mm}$.}
    \label{fig:ratio}
\end{figure}

Moreover, the FI framework allows us to estimate the overall contribution of polarimetric information. Due to the additive nature of the FI, by considering only selected de-rotated MM elements, we can compare the minimum achievable statistical error via the CRB. Based on this approach, Fig.~\ref{fig:ratio} shows the ratios of the standard deviations obtained when using only $\widetilde{M}_{11}$ (brown line), $\widetilde{M}_{11}$, $\widetilde{M}_{12}$, $\widetilde{M}_{22}$ (blue line) and $\widetilde{M}_{11}$, $\widetilde{M}_{12}$, $\widetilde{M}_{22}$, $\widetilde{M}_{33}$, $\widetilde{M}_{34}$ (red line), relative to the full set of de-rotated MM elements. The figure illustrates the impact of approaches based on measuring only a limited number of polarization states rather than reconstructing the full MM \cite{Stefanov:23,falconet2008analysis,chae2025machine}. As can be seen, using a truncated MM may be reasonable for certain samples. However, this comparison should be regarded only as an illustration for an ideal experiment, since it neglects the possible presence of systematic errors in the setups, which can significantly affect the relative contribution of different MM elements.

The presented setups and data processing framework allow not only to quantitative evaluate experimental setups but also to estimate in a controlled way the best polarimetric parameters for the identification of specific features of the scattering media. This is particularly relevant for biological tissues for which the modeling anisotropic scattering is very difficult (like tumorous brain tissue) due to the tissues complexity at various scales \cite{menzel2020toward}. Using a simplified model of the tissues in a MC simulation and implementing it in corresponding phantoms~\cite{chue2019optical} would lead to a better understanding of the interaction of tissues with polarimetric light.

The further development of the proposed approach may also provide insights for interpreting the responses of different machine-learning models due to the explicit formulation of the optimization problem used for sample characterization.

\section*{Funding}
Schweizerischer Nationalfonds zur Förderung der Wissenschaftlichen Forschung (200021\_212872).

\section*{Acknowledgment}
This work is supported by the Swiss National Science Foundation (Grant No. 200021\_212872).

\section*{Disclosures}
The authors declare no conflicts of interest.

\section*{Data availability}

The data that support the findings of this study are openly available in the Bern Open Research Information System (BORIS) at https://doi.org/10.48620/96911

\appendix

\section{Maximum likelihood method}\label{append}

A maximum likelihood method is based on the solving of the optimization problem for searching of parameters, that maximize a log-likelihood function. Assuming the noise is Gaussian, we can write the log-likelihood as
\begin{equation}
    \mathfrak{l}(\zeta)=-\frac{1}{2\sigma^2}\sum_s Tr\left[ (I_s-\beta_s A M_s W)\cdot (I_s-\beta_s A M_s W)^T\right], \label{log-likelihood}
\end{equation}
where $\sigma$ represents the standard deviation of the Gaussian noise, $s$ denotes the measurement number, $\beta_s$ stands for the effective transmittance coefficient, and $M_s$ is the MM corresponding to the $s$-th measurement.

Expressing transmittance coefficients $\beta_s$ from Eq.~\eqref{log-likelihood} in form
\begin{eqnarray}\label{coef_MLM}
    \beta_s=\frac{\sum\limits_{i_s} Tr\left[(A M_{i_s} W)\cdot I_{i_s}^T \right]}{\sum\limits_{i_s} Tr\left[(A M_{i_s} W)\cdot (A M_{i_s} W)^T \right]},
\end{eqnarray}
allows to rewrite expression for maximization to
\begin{equation}\label{MLM}
    \mathfrak{l}(\zeta)=\sum_s\frac{\left(\sum\limits_{i_s} Tr\left[(A M_{i_s} W)\cdot I_{i_s}^T \right]\right)^2}{\sum\limits_{i_s} Tr\left[(A M_{i_s} W)\cdot (A M_{i_s} W)^T \right]},
\end{equation}
where we have $s$ sets of $i_s\geq1$ measurements with the equal staff.

\bibliography{sample}
\end{document}